\documentstyle[manuscript,aps]{revtex}
\begin{document}
\draft
\title{On the gravitational energy of the Bonnor spacetime}
\author{Janusz Garecki}
\address{Institute of Physics, University of Szczecin, Wielkopolska 15; 70-451
Szczecin, POLAND\footnote{e-mail:garecki@sus.univ.szczecin.pl}}
\date{\today}
\maketitle
\begin{abstract}
In the paper we consider the gravitational energy and its flux in the Bonnor
spacetime. We construct some non-local expressions from the Einstein
canonical energy-momentum pseudotensor of the gravitational field which
show that the gravitational energy and its flux in this spacetime are
different from zero and do not vanish even outside of the material
source (a stationary beam of the null dust) of this spacetime.
\end{abstract}
\section{Introduction}
In the paper [1] W.B. Bonnor gave the solution to the Einstein equations
which describes a stationary beam of light (stricly speaking --- a
stationary beam of null dust). The line element $ds^2$ for this solution
reads $(G=c=1)$
\begin{equation}
ds^2 = \bigl(1+m\bigr)dt^2 - 2mdtdz -\bigl(1-m\bigr)dz^2 - dx^2-dy^2,
\end{equation}
where
\begin{equation}
\Delta m:={\partial^2m\over\partial x^2} + {\partial^2m\over\partial
y^2} = 16\pi\rho,
\end{equation}
and $\rho = \mu(u^0)^2> 0$.

$\mu\geq 0$ is the rest density of the null dust and $:=$ means ``by
definition''.

In the paper [2] were calculated the total energy-momentum
``densities'', matter and gravity, by using standard energy-momentum
complexes of Einstein, Landau-Lifschitz, Papapetrou and Weinberg. All
the calculations were performed after transforming the line element (1)
to the {\it Kerr-Schild form} (see, e.g., [2])

It was shown in [2] that the used four energy-momentum complexes give
compatible results in the case and ``localize'' total energy-momentum
``densities'' in the Kerr-Schild coordinates to the domains occupied by
sources of the gravitational field only, i.e., to the domains in which
energy-momentum tensor of the null dust does not vanish. Vacuum regions
of the Bonnor spacetime give no energy-momentum contribution, i.e., the
gravitational energy-momentum pseudotensors of Einstein,
Landau-Lifschitz, Papapetrou and Weinberg {\it globally vanish} in the
Kerr-Schild coordinates outside of the stationary beam of the considered
null dust.

Then, the analogical calculations were repeted in [3] but in Bonnor's
coordinates $x^0 = t,~x^1 = z, ~x^2 = x, ~x^3 = y$. The results were the
same as the results obtained earlier in [2]: in the used coordinates
$(t,z,x,y)$ the total energy and momentum ``densities'' were confined to
the regions of non-vanishing energy-momentum tensor of the null dust,
i.e., again the vacuum regions of the Bonnor spacetime gave no
energy-momentum contribution.

Despite the fact that their results are {\it coordinate-dependent}, the
authors of the papers [2,3] conclude that their results sustain the
hypothesis which states that the energy and momentum in general
relativity ({\bf GR}) are confined to the regions of non-vanishing
energy-momentum tensor of matter.\footnote{This old hypothesis was
recently restored, e.g., by F. Cooperstock} From this hypothesis it
follows, e.g., that the gravitational waves carry no energy-momentum.

In the following we will show that the conclusion of such a kind {\it
cannot be correct}.

\section{The gravitational energy and its flux in the Bonnor spacetime outside of the
beam}
At first, let us observe that one can write the line element (1)-(2) in
the form
\begin{equation}
ds^2 = dt^2 + m\bigl(dt - dz\bigr)^2 - dz^2 - dx^2 -dy^2,
\end{equation}
where
\begin{equation}
\Delta m = 16\pi\rho\geq 0.
\end{equation}
Then, after introducing new variables $u = t-z,~v = t+z$ we get
\begin{equation}
ds^2 = mdu^2 + dudv - dx^2 - dy^2,
\end{equation}
\begin{equation}
\Delta m = 16\pi\rho\geq 0.
\end{equation}

Finally, by introducing $U = {u\over\sqrt{2}},~V = {v\over\sqrt{2}},~X
=x,~Y=y$ one gets the line element (1)-(2) in the form
\begin{equation}
ds^2 = 2mdU^2 + 2 dUdV - dX^2 - dY^2,
\end{equation}
where
\begin{equation}
\Delta m = {\partial^2m\over\partial X^2} + {\partial^2m\over\partial
Y^2} = 16\pi\rho.
\end{equation}
In vacuum, i.e., outside of the stationary beam of the null dust which
is the source of the Bonnor spacetime, we obtain from (7)-(8)
\begin{equation}
ds^2 = 2 m(X,Y)dU^2 + 2 dUdV - dX^2 - dY^2,
\end{equation}
\begin{equation}
\Delta m = {\partial^2m\over\partial X^2} + {\partial^2m\over\partial
Y^2} = 0,
\end{equation}
i.e., we obtain special {\it plane-fronted gravitational wave with
parallel rays} (p-p wave, see, e.g., [4]).

In an old our paper [5] it was shown that in the coordinates $(U,V,X,Y)$
the canonical Einstein's gravitational energy-momentum pseudotensor $_E t_i^{~k}$
globally vanishes for the p-p waves giving zero gravitational ``energy
density'' and no gravitational energy flux, in agreement with the results of
the papers [2,3]. But if one transforms the line element (9)-(10) to the
coordinates used by MTW [6], then one will obtain non-vanishing
pseudotensor $_E t_i^{~k}$ and nonzero (negative) ``energy density'' and
its non-zero flux (see, e.g., [5]).

So, as it is commonly known, the results obtained by using of the
energy-momentum pseudotensors of the gravitational field (and complexes)
are {\it coordinate - dependent} and one {\it must not use} them for supporting the
hypothesis about ``localization of the gravitational energy to the regions of
the non-vanishing energy-momentum tensor of the matter and all
non-gravitational fields''.

The energy-momentum complexes, matter and gravitation, and in
consequence the gravitational energy-momentum pseudotensors, can only be
reasonably use in a case of very precisely defined asymptotically flat,
in null or in spatial infinities, spacetime giving global (or
integral) energy and momentum. But they {\it cannot have} physical
meaning to any local analysis of the gravitational field.

In order to do a reasonable and covariant local analysis of the
gravitational field one should use {\it covariant expressions}
constructed from the curvature tensor, e.g., one should use the {\it
canonical superenergy tensor} $_g S_i^{~k}(P;v^l)$ of the gravitational
field (and canonical angular supermomentum tensor).

The tensor of such a kind was introduced in series of the our papers
[7-12].  In vacuum, the canonical superenergy tensor of the
gravitational field has the following form
\begin{eqnarray}
_g S_i^{~k}(P;v^l)&=& {8\alpha\over 9}\bigl(2v^av^b -
g^{ab}\bigr)\biggl[R^{klm}_{~~~(a\vert}{}R_{i(lm)\vert b)}\nonumber\\
&-& {1\over 2}\delta_i^k {}R^{lmn}_{~~~(a\vert}{}R_{l(mn)\vert
b)}\biggr],
\end{eqnarray}
where $\alpha = {1\over 16\pi}$ and Latin indices run over the values
$0,1,2,3$.

Here $v^av_a = 1$ means a 4-velocity of an observer {\bf O} which is
studying gravitational field and round brackets denote symmetrization.
The indices inside vertical lines, e.g., $(a\vert bc\vert d)$ are
excluded from symmetrization.

The tensor (11) was obtained as a result of a special averaging of the
differences $_E t_i^{~k}(y) - {_E t_i^{~k}}(P)$ in a Riemann normal
coordinates {\bf NC(P)} introduced in a sufficiently small vicinity of
an (arbitrary) point {\bf P} (For details--see, e.g., [7-12]). So, it
is some kind of a non-local construction obtained from the Einstein
canonical energy-momentum pseudotensor $_E t_i^{~k}$.

The fiducial observer {\bf O} is at rest in this {\bf NC(P)}. In
Bonnor's coordinates $(t,z,x,y)$ the four-velocity $\vec v$ of this
observer has the following components\footnote{At every point {\bf P} we
have chosen the unit timelike vector of the {\bf NC(P)} to be proportional
to the timelike vector of the holonomic frame at the point determined by the
Bonnor's coordinates.} 
\begin{equation}
v^i = {\delta^i_0\over\sqrt{g_{00}}} = {\delta^i_0\over\sqrt{1+m}},
\Rightarrow v_i = {g_{i0}\over\sqrt{1+m}}.
\end{equation}

In the paper [5] were calculated the components $_g S_i^{~k}(P;v^l)$ of
the canonical superenergy tensor (11) for a gravitational p-p wave in
{\it the null coreper}
\begin{equation}
\vartheta^0 = mdU+dV,~\vartheta^1 = dU,~\vartheta^2 = dX,~\vartheta^3 =
dY
\end{equation}
determined by the line element (9)-(10). There was obtained that the
only one component
\begin{equation}
_g S_0^{~1} = {16\alpha\over 9}(v^1)^2\bigl[(m_{xx})^2 + 2(m_{xy})^2 +
(m_{yy})^2\bigr]
\end{equation}
of the $_g S_i^{~k}(P;v^l)$ is different from zero in the coreper (13).

Here and in the following $m_{xx} = {\partial^2m\over\partial
x^2},~m_{xy} = {\partial^2m\over\partial x\partial y},~m_{yy} =
{\partial^2m\over\partial y^2}$.

By using (14), one can easily calculate the canonical superenergy
density $^g\epsilon_s$ of the p-p wave (9)-(10) in the null coreper
\begin{eqnarray}
^g\epsilon_s &:=& _g S_i^{~k}v^iv_k {\dot =} _g S_1^{~0} (v^1)^2 {\dot
=} _g S_1^{~0}(v_0)^2\nonumber\\
&=& {4\alpha\over 9(m+1)^2}\biggl[(m_{xx})^2 + 2(m_{xy})^2 +
(m_{yy})^2\biggr] > 0.
\end{eqnarray}
The sign ${\dot =}$ means that an equation is valid only in some special
coordinates or frames.

The expression (15), like the expression (14), is {\it
positive-definite} and {\it it does not vanish} in any coordinates or
frames.

Also the spatial Poynting's supervector
\begin{equation}
_g P^i :=\bigl(\delta^i_k - v^iv_k\bigr)_g S_l^{~k}v^l
\end{equation}
does not vanish in the case and it has the two non-vanishing components
$P^0,~P^1$ in the null coreper (13):
\begin{equation}
P^0{\dot =} {8\alpha\over 9\sqrt{2(m+1)^3}}\bigl[(m_{xx})^2 + 2
(m_{xy})^2 + (m_{yy})^2\bigr],
\end{equation}
\begin{equation}
P^1{\dot =} {-8\alpha\over 9(m+1)\sqrt{2(m+1)^3}}\bigl[(m_{xx})^2 +
2(m_{xy})^2 + (m_{yy})^2\bigr] = {-P^0\over(m+1)}.
\end{equation}

So, we have for Bonnor spacetime (1)-(2) the {\it non-vanishing} gravitational
superenergy density and {\it non-vanishing} gravitational superenergy
flux even outside of the sources of this spacetime, i.e., outside of the
domains in which $T_i^k \not= 0$.

As we have used tensorial expressions only in our analysis of the Bonnor
spacetime, the our results are valid in any coordinates or frames. Thus,
one can conclude that in the case of the Bonnor spacetime one has a {\it
positive-definite} gravitational superenergy density and {\it
non-vanishing} gravitational superenergy flux in vacuum, i.e., outside
of the stationary beam of the null dust which is the source of this
spacetime.\footnote{ Of course, the same conclusion is also true inside
of the beam. However, in this domain the suitable expression on $_g
S_i^{~k}(P;v^l)$ is slightly modified by additional terms which depend
on the matter tensor $T_i^k$. On the other hand, the canonical
superenergy tensor of matter $_m S_i^{~k}(P;v^l)$ defined in [7-12] is
confined to the domains occupied by matter, i.e., it is confined to the
beam only.}

But this means that the free gravitational field in the Bonnor spacetime
for which $R_{iklm}\not= 0$ {\it also possesses} its own gravitational
(relative) energy-density and the {\it non-zero} of this (relative)
gravitational energy flux. It is easily seen from the following
considerations.

Let us consider an observer {\bf O} which is studying gravitational
field. His world-line is $x^a = x^a(s)$, and $\vec v: ~v^a = {dx^a\over
ds}$ represents his four-velocity. At any point {\bf P} of the
world-line one can define an instantaneous, local 3-space of the observer
{\bf O} orthogonal to $\vec v$. This instantaneous 3-space has the following
interior proper Riemannian metric
\begin{equation}
\gamma_{ab} = v_av_b - g_{ab}{\dot =} \biggl({g_{0a}g_{0b}\over g_{00}}
- g_{ab}\biggr){\dot =}\gamma_{\alpha\beta} {\dot =}(-)g_{\alpha\beta},
\end{equation}
where the Greek indices run over the values $1,2,3$ (see, e.g., [13]).

Then, by using the gravitational superenergy density $^g\epsilon_s$ and
its flux $_g P^i$, one can easily construct in such instantaneous local
3-space of the observer {\bf O}, e.g., the following non-local expressions
which have proper dimensions of the energy density and its flux
\begin{eqnarray}
\epsilon_{en} &:=&\oint\limits_{S_2}{^g\epsilon_s(P)
d^2S}\nonumber\\
&\approx& ^g\epsilon_s(P)\oint\limits_{S_2}{d^2S},
\end{eqnarray}
\begin{eqnarray}
P^i&:=& \oint\limits_{S_2}{_g P^i(P)d^2S}\nonumber\\
&\approx&{_g P^i(P)}\oint\limits_{S_2}{d^2S}.
\end{eqnarray}

Here $S_2$ means a small sphere $\gamma_{\alpha\beta}x^{\alpha}x^{\beta}
= R^2$ in the instantaneous local 3-space
of the observer {\bf O}.
\begin{equation}
d^2S =\sqrt{\gamma_{\alpha\beta} \sigma^{\alpha}\sigma^{\beta}},
\end{equation}
where
\begin{equation}
\sigma^{\alpha}= {1\over
2}\gamma^{\alpha\beta}{}\epsilon_{\beta\gamma\delta }dx^{\gamma}\wedge
dx^{\delta}, 
\end{equation}
and $\gamma^{\alpha\beta}$ means the inverse metric to the interior
metric $\gamma_{\alpha\beta}$.

$\epsilon_{\alpha\beta\gamma}$ is the 3-dimensional Levi-Civita
pseudotensor established by the condition 
\begin{equation}
\epsilon_{123} = \sqrt{\gamma},
\end{equation}
where $\gamma := det(\gamma_{\alpha\beta})$. 
The expressions (20)-(21) give us the {\it relative gravitational energy
density} and its flux for an observer {\bf O} in his instantaneous
3-space orthogonal to $\vec v$.\footnote{The fact that (20)-(21) give us some
kind of the relative quantities (with respect {\bf P}) is seen from the method
of the construction of the canonical superenergy tensors (see, e.g.,
[7-12]).} So, the values of these quantities $\epsilon_{en}$ and
$P^i$ depend on radius $R$ of the sphere $S_2$ also. 
In order to get unique $\epsilon_{en}$ and $P^i$ one can take as the
sphere $S_2$, e.g., the smallest classical sphere admitted by the loops
quantum gravity.

This quantum theory of gravity tells us (see, e.g., [14-23]) that one can say
about continuous classical differential geometry already just a few orders of
magnitude above the Planck scale, e.g., for distances $L\approx
100 L_P\approx 10^{-33}m$ . So, one can take the radius
$R$ of the smallest classical  sphere in an instant 3-space of the
observer {\bf O}, orthogonal to $\vec v$, to be of order $ R\approx 100 L_P
= 100\cdot\sqrt{{G\hbar\over c^3}} \approx 10^{-33}m$. In such a case,
when calculating the integrals (20)-(21) one can assume that their integrands
are constant during integration and evaluate these integrals in the way as it
was already done in (20)-(21).

As we could have seen from this paper the two quantities $\epsilon_{en}$
and $P^i$ {\it do not vanish} in the Bonnor spacetime for every {\bf O};
especially, they {\it do not vanish} outside of the beam which is the
source of this spacetime. In consequence, one can conclude that it is easy to
attribute to the gravitational field in Bonnor spacetime the
positive-definite (although relative) energy-density and its non-zero flux
by using, e.g., our canonical superenergy tensor of this field.

When we only use gravitational pseudotensors, then this important fact is
camouflaged in some coordinates, like Kerr-Schild coordinates or Bonnor
coordinates, by energy and momentum of the inertial forces field (with
$R_{iklm} = 0$) which simply cancels (outside of the beam) with energy and
momentum of the real gravitational field (with $R_{iklm}\not = 0$).

Thus, the conclusions of the authors of the papers [2,3] {\it are
incorrect} as they resulted from coordinate-dependent pseudotensorial
expressions. In order to get the correct information about gravitational
energy and momentum (and also about gravitational angular momentum, see,
e.g., [11,12]) one must use covariant expressions which depend on the
curvature tensor, like our canonical gravitational superenergy tensor
$_g S_i^{~k}(P;v^l)$ (and like our gravitational angular supermomentum
tensors, see, e.g., [11,12]).

\end{document}